\documentstyle[seceq,epsf]{ptptex}

\title{Quantum Creation of the Randall-Sundrum Bubble
}

\author{
Hirotaka {\sc Ochiai}$^1$, Daisuke {\sc Ida}$^{2}$ and 
Tetsuya {\sc Shiromizu}$^{3}$
}

\inst{
$^1$Department of Physics, The University of Tokyo, Tokyo 113-0033 
\\
$^{2,3}$ Research Center for the Early Universe (RESCEU),
The University of Tokyo, Tokyo 113-0033 
}

\recdate{
}

\abst{
We investigate the semiclassical instability of the Randall-Sundrum 
brane world. We carefully analyze the bubble solution with the 
Randall-Sundrum background, which expresses the decay of the brane world. 
We evaluate the decay probability following the Euclidean 
path integral approach to quantum gravity. Since a bubble rapidly 
expands after the nucleation,  the entire spacetime will be occupied
by such bubbles.
}

\begin{document}

\maketitle

\section{Introduction}

In the nonperturbative approach to string theory, it is becoming accepted 
that the standard model particles are confined to branes. \cite{HW}
This idea has led to the so-called brane-world scenario of the universe.
The simplest models describing this scenario have been proposed by Randall and Sundrum. \cite{RSI,RSII}
Therein, the four-dimensional Minkowski brane 
is located at the boundary of the five-dimensional 
anti-de Sitter (adS) space-time.
The first Randall-Sundrum model (RS1) was motivated by the hierarchy problem
and consists of two 3-branes~\cite{RSI}. 
The second Randall-Sundrum model (RS2) consists of a single 3-brane.~\cite{RSII}
It has been shown that four-dimensional gravity,
not five-dimensional gravity,
is recovered at low energy scales
on the branes.~\cite{Tama,Tess}
This represents a new type of dimensional reduction that is an alternative
to the Kaluza-Klein compactification.
In this way, possibility of realizing noncompact extra dimensions arises. 
The RS models can describe standard cosmology at low energy 
scales. There are indeed exact solutions describing the homogeneous and 
isotropic expanding universe. \cite{Cosmos} 
Moreover, the RS brane world has 
comprehensive features, e.g., 
the adS/CFT interpretation. \cite{adSCFT} 

Although RS models have had great success, 
there are still fundamental questions regarding 
the stability. There are mainly two problems with RS1 models. 
One involves
the radius stabilizations: In order to recover the correct four-dimensional 
gravity, we must 
assume that the distance between branes is stabilized. \cite{Tama2} 
A toy model for the stabilization problem has been proposed by  
Goldberger and Wise. \cite{gw}

The stability problem that we consider in the 
present paper concerns a quantum process. 
In the unified Kaluza-Klein theory,
it is well known that the Kaluza-Klein vacuum is unstable
with respect to the decay channel
to the Kaluza-Klein bubble space-time.~\cite{KKB,Gary}
It may be thought that there is a similar instability of the RS model.
This was first pointed out  in Refs.~\cite{Shinkai} and ~\cite{Horava}.
The discovery of an explicit example 
describing a sort of Kaluza-Klein bubble in the RS1 model
(the RS bubble) was reported 
in a previous paper.\cite{rsb} 
(A somewhat relevant solution is presented 
in Ref. \cite{Ruth})
In order to make the RS model feasible,
this decay channel 
must be suppressed in some way.
The purpose of this paper is to
estimate the transition rate of the RS vacuum to the RS bubble in 
the framework of the Euclidean path integral procedure. 

The remainder of this paper is organized as follows. 
In \S \ref{II}, we briefly review the RS bubble space-time.
In \S \ref{III}, we calculate the transition probability of 
the RS vacuum to the RS bubble spacetime.
In \S \ref{IV} we summarize the present work. For simplicity, we will set 
the five-dimensional gravitational scale to unity: $G_5=\kappa_5^2=1$.

\section{Randall-Sundrum bubble}\label{II}

Here we briefly review the Randall-Sundrum bubble introduced in a 
previous paper.~\cite{rsb} 
The RS model of a two brane system (RS1)~\cite{RSI} 
is given by the metric
%
\begin{equation}
g=dy^2+e^{-2y/\ell}\eta_{\mu\nu}dx^\mu dx^\nu,\label{RS}
\end{equation}
%
where $\eta_{\mu\nu}$ 
is the four-dimensional Minkowski metric and $0\le y\le y_0$.
The metric (\ref{RS}) is that of the five-dimensional adS space,
and positive and negative tension branes are located at $y=0$ and $y=y_0$,
respectively.
The tension $\lambda$ of the ($\pm$)-branes are given by $\pm 6/\ell$,
and the extrinsic curvatures on the branes
are given by
%
\begin{equation}
K_{\mu\nu}:=\frac{1}{2} \mbox \pounds_n h_{\mu\nu}
=\mp\frac{1}{\ell} h_{\mu\nu},
\label{junction}
\end{equation}
%
where $n=\pm\partial_y$ is the outward unit vector normal to the boundary
and $h_{\mu\nu}=e^{-2y/\ell}\eta_{\mu\nu}$. 

If the four-dimensional metric $\eta_{\mu\nu}$ is replaced by a Ricci-flat
metric $q_{\mu\nu}$, then Eq.~(\ref{RS}) represents a more generic Einstein metric.
Let us write the brane-metric in the form
%
\begin{equation}
q=-r^2d\tau^2+dr^2+r^2\cosh^2\tau d\Omega_2{}^2,\label{Rindler}
\end{equation}
%
where $d\Omega_2{}^2$ denotes the standard metric of the unit two-sphere.
The metric (\ref{Rindler}) represents the spherical 
Rindler space, which is locally flat
but geodesically incomplete at the null hypersurface $r=0$ (Rindler horizon).
Each $r={\rm constant}$ hypersurface corresponds to a world sphere
in a uniformly accelerating expansion.
We here consider another generalization of Eq.~(\ref{RS}) with the same
asymptotic behavior as Eq.~(\ref{Rindler}) on the brane.
This is given by
\footnote{Carrying out the signature change of $\Lambda$ and 
the double Wick rotation [$\Lambda \to -\Lambda, 
y \to it$ and $\tau \to i(\theta +\pi/2)$], it is found that 
this metric describes the five-dimensional Schwarzshild-deSitter space-time 
in isotropic coordinates.
For investigation of four and higher
dimensional cases, see Refs.~\cite{dSCFT} and ~\cite{Iso}.}
%
\begin{eqnarray}
g=\left[\frac{1-(\rho_*/ar)^2}{1+(\rho_*/ar)^2}\right]^2dy^2
+a^2\left[1+\left(\frac{\rho_*}{ar}\right)^2\right]^2(-r^2d\tau^2
+dr^2+r^2\cosh^2\tau d\Omega_2{}^2),\nonumber \\ 
\label{exp}
\end{eqnarray}
%
where
$\rho_*>0$ is a constant, $a:=e^{-y/\ell}$ and $0\le y\le y_0$.
The positive and negative tension branes are located at $y=0$ and $y=y_0$,
respectively. 
The metric (\ref{exp}) solves the five-dimensional Einstein equation 
with a negative cosmological term, and Eq.~(\ref{junction}) is also 
satisfied on every $y={\rm [constant]}$ hypersurface. The coordinate 
system used here is inappropriate at $ar=\rho_*$, However, this is 
only a coordinate singularity, as shown below. 

\begin{figure}[t]
\vspace*{-5mm}
\begin{center}
\epsfxsize=3.0in
\epsffile{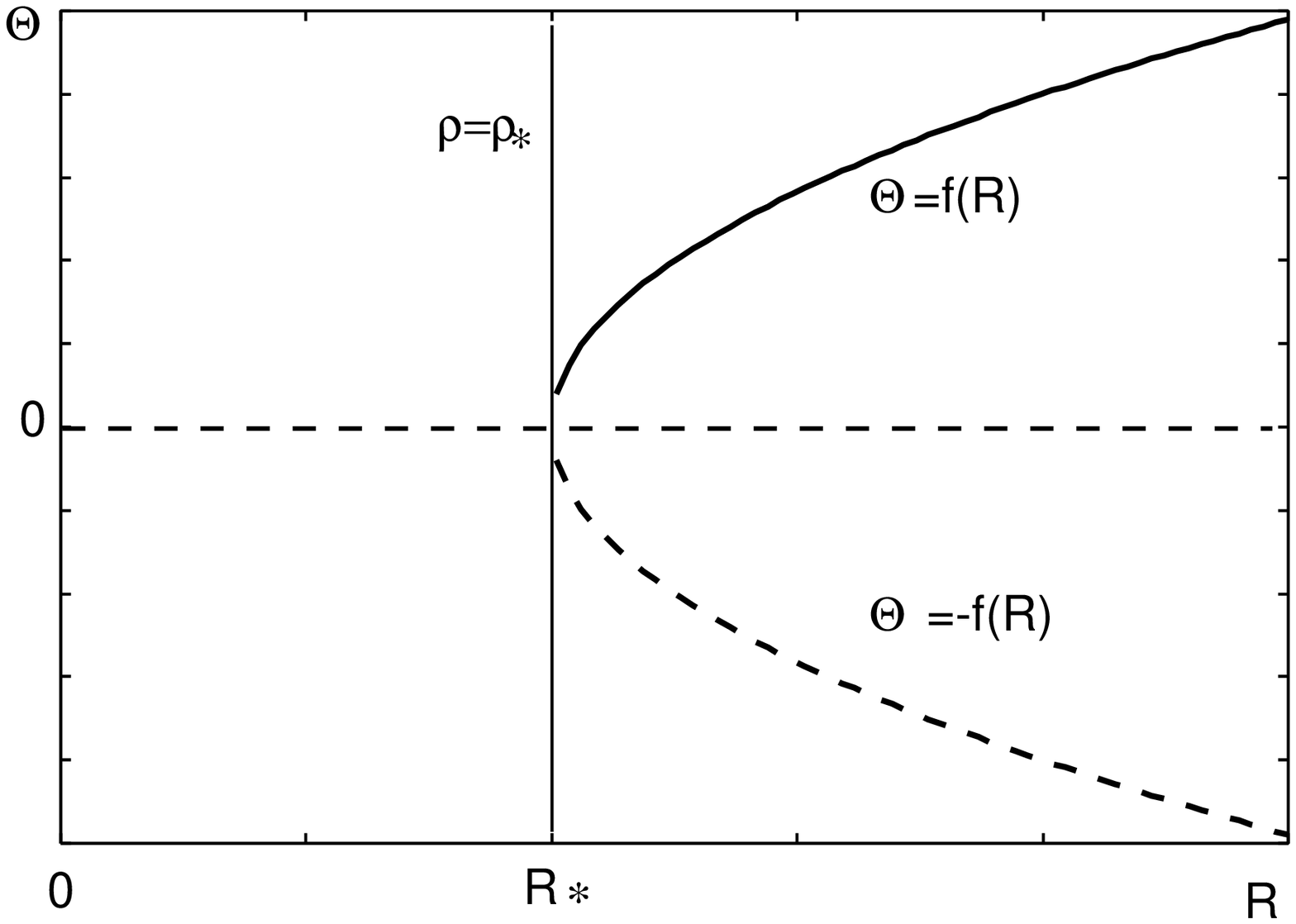}
\end{center}
\caption{The location of a ($+$)-brane ($y=0$)
in the ($R$, $\Theta$)-plane.}
\label{fig1}
\end{figure}

\begin{figure}[t]
\vspace*{-5mm}
\begin{center}
\epsfxsize=3.0in
\epsffile{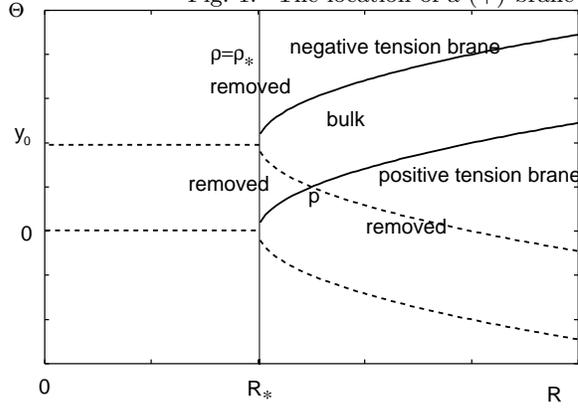}
\end{center}
\caption{The location of ($\pm$)-branes ($y=0$, $y_0$)
in the ($R$, $\Theta$)-plane. $p$ denotes the intersection surface of two 
3-branes.}
\label{fig2}
\end{figure}

{}From this point, we construct RS bubble space-times in the global sense. 
Then we investigate the global structure of the RS bubble space-times.
The metric (\ref{exp}) is obtained by analytic continuation of 
the five-dimensional adS-Schwarzschild space-time, whose metric 
has the form
%
\begin{eqnarray}
& & g=-F(R)dT^2+F(R)^{-1}dR^2+R^2 (d\chi^2+\sin^2\chi d\Omega_2{}^2),\label{schads}\\
& & F(R)=1-\left(\frac{R_*}{R}\right)^2+\left(\frac{R}{\ell}\right)^2.
\end{eqnarray}
%
This metric can be analytically continued on the totally geodesic surfaces
$T=0$ and $\chi=\pi/2$ by the replacement of the coordinates
%
\begin{equation}
T\mapsto i\Theta,~~~\chi\mapsto\frac{\pi}{2}+i\tau.
\end{equation}
%
Then the metric becomes
%
\begin{equation}
g=F(R)d\Theta^2+F(R)^{-1}dR^2+R^2 (-d\tau^2+\cosh^2\tau d\Omega_2{}^2),
\label{newschads}\\
\end{equation}
%
which represents the straightforward generalization of the Kaluza-Klein 
bubble in the local sense.
To arrive at the brane-world metric (\ref{exp}), we consider the
coordinate transformation from ($\Theta$, $R$) to ($y$, $t$) defined by

\begin{eqnarray}
R&=&ar\left[1+\left(\frac{\rho_*}{ar}\right)^2\right],\\
\Theta&=&y+\frac{1}{\ell}\int^R_{R_*}\frac{R}{F(R)}
\left(1-\frac{R_*{}^2}{R^2}\right)^{-1/2}dR,
\end{eqnarray}
where $\rho_*=R_*/2$, with
$-\infty < y < \infty$, and $ar>\rho_*$.
This chart covers the region $R>R_*$ in the $(\Theta, R)$ coordinate 
system. 
The $y=0$ surface is geodesically incomplete at 
$r=\rho_*$ [$(\Theta,R)=(0,R_*)$]. 
It can easily be made geodesically 
complete by reflecting with respect to the surface $\Theta=0$.
If the $y=0$ surface is given by $B_+$: $\{\Theta=f(R)\}$, then the 
reflected surface $B_-$: $\{\Theta=-f(R)\}$ smoothly continues to $B_+$ 
at $R=R_*$ (see Fig.~\ref{fig1}).
Similarly, the negative tension brane at $y=y_0$ can be obtained by
$\bar B=\bar B_+\cup \bar B_-$, where 
$\bar B_\pm$: $\{\Theta=\pm f(R)+y_0\}$ (see Fig.~\ref{fig2}).
Cutting and gluing the RS bubble spacetime with two 3-branes
is obtained as in Fig. 2.

\begin{figure}[t]
\vspace*{-5mm}
\begin{center}
\epsfxsize=3.0in
\epsffile{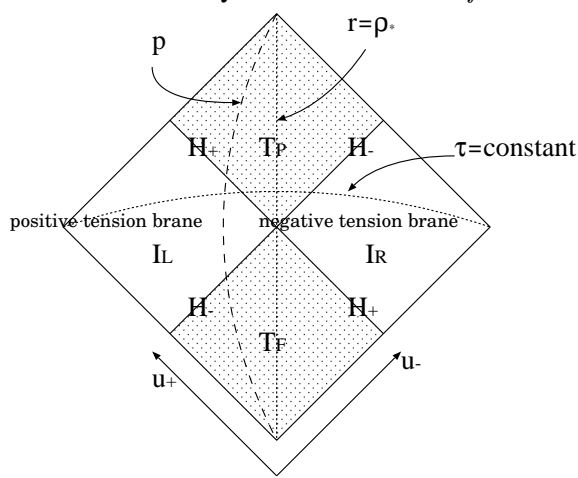}
\end{center}
\caption{The Penrose diagram for the induced geometry of the brane
corresponding to
the situation in Fig. 2.
The dashed curve denotes the surface $p$ connecting the two branes.}
\label{fig3}
\end{figure}

Finally, consider the induced geometry of the brane.
It can be shown that the induced metric, $h$, is given by
%
\begin{equation}
h=\left[1+\left(\frac{\rho_*}{r}\right)^2\right]^2(-r^2d\tau^2+dr^2+r^2\cosh^2\tau
d\Omega_2{}^2).
\label{branemetric}
\end{equation}
%
The coordinate $r$ now ranges over all positive values,
where the region $r>\rho_*$ 
corresponds to $B_+$ and $0<r<\rho_*$ to $B_-$ [note that the metric (\ref{branemetric}) 
is invariant under $r\mapsto\rho_*^2/r$].
Next, let us introduce the null coordinates 
$u_\pm=\tau\mp\ln (r/\rho_*)$.
In terms of these, the metric (\ref{branemetric}) becomes
%
\begin{eqnarray}
h=-\rho_*{}^2e^{u_++u_-}\left(e^{-u_+}+e^{-u_-}\right)^2du_+du_-
+{\cal R}(u_+,u_-)^2d\Omega_2{}^2,
\end{eqnarray}
%
where
%
\begin{equation}
{\cal R}(u_+,u_-)=\frac{\rho_*}{2}(1+e^{u_++u_-})(e^{-u_+}+e^{-u_-}).
\end{equation}
%
Then, the expansion rates of the outgoing and ingoing spherical rays are 
given by
%
\begin{equation}
\theta_\pm:=\frac{\partial\ln {\cal R}}{\partial u_\mp}=\pm\frac{e^{\pm u_\mp}-e^{\mp u_\mp}}{(1+e^{u_++u_-})(e^{-u_+}+e^{-u_-})},
\end{equation}
%
respectively.
There are null hypersurfaces $H_+$ and $H_-$,
\begin{equation}
H_\pm: u_\mp=0,
\end{equation}
 on which $\theta_+$ and $\theta_-$ vanish, respectively.
The brane $(B,h)$ is divided by $H_+$ and $H_-$ into
four regions:
\begin{enumerate}
\item $I_R$ (right asymptotic region):\\
$u_+<0$, $u_->0$ [$(\theta_+,\theta_-)=(+,-)$].
\item $I_L$ (left asymptotic region):\\
$u_+>0$, $u_-<0$ [$(\theta_+,\theta_-)=(-,+)$].
\item $T_P$ (past trapped region):\\
$u_+>0$, $u_->0$ [$(\theta_+,\theta_-)=(+,+)$].
\item $T_F$ (future trapped region):\\
$u_+<0$, $u_-<0$ [$(\theta_+,\theta_-)=(-,-)$].
\end{enumerate}
The Penrose diagram describing to
the situation is depicted in Fig.~\ref{fig3}.

Each $\tau={\rm constant}$ hypersurface has an Einstein-Rosen 
bridge around $r=\rho_*$. Thus, both the bulk and the brane have
non-trivial topology
(simply connected but with non-vanishing second Betti number),
which represents the creation of a sort of bubble.
However it is not possible to traverse from one side to the other:
Once $T_P$ is entired, it is impossible to leave.
Thus, the region $T_P$ is a kind of black hole,
though it is very different from what we know of black holes.
In particular, the total
gravitational energy vanishes, which indicates that the vacuum in the RS1 model
might decay by creating RS bubbles semi-classically, as we will discuss 
in the next section. This strange structure results from the fact that 
there is an effective negative energy distribution on the brane, 
which is due to the electric part of the five-dimensional Weyl 
tensor.\cite{rsb}  

As mensioned above, in the RS1 model,
the RS vacuum might decay through the semi-classical creation
of the RS bubbles.
The creation of a bubble implies a connection of two branes
through a topology-changing process of
the bulk and the brane.
To be a good model of the universe,
there should be a stabilization against this kind of instability
in the RS1 model.
In the next section we
estimate the decay probability to an RS bubble in the RS1 model.

\section{Quantum creation of the Randall-Sundrum bubble}\label{III}
\subsection{Decay rate}

We now discuss the decay of the RS vacuum to RS bubble spacetime.
The corresponding Euclidean 
solution is obtained by the Wick rotation
$\tau \to i \tau_E +\pi/ 2$ of the metric of 
Eq. (\ref{exp}):
%
\begin{eqnarray}
g & = & \Bigl[\frac{1-(\rho_*/ar)^2}{1+(\rho_*/ar)^2}  \Bigr]^2dy^2
+a^2\Bigl[1+\Bigl( \frac{\rho_*}{ar}\Bigr)^2  \Bigr]^2 \Bigl(dr^2 
 +r^2d\Omega_3^2 \Bigr),
\end{eqnarray}
where $d\Omega_3^2=d\tau_E^2+{\rm sin}^2\tau_E d\Omega_2^2$ is the standard metric of $S^3$.

{}From this point we estimate the nucleation probability of an RS bubble. 
Since we are interested in the transition rate of the RS vacuum to an RS bubble, 
we can use the following formula for the decay probability:\cite{Gen} 
%
\begin{eqnarray} 
P \sim e^{-(S_E-S_E^{(0)})}=:e^{-[S_E]},
\end{eqnarray}
%
where $S_E$ and $S_E^{(0)}$ are the Euclidean action of the RS bubble and 
the RS vacuum, respectively. 
The Euclidean action $S_E^{(0)}$ is evaluated in the RS vacuum with the metric
%
\begin{eqnarray}
g&=&dy^2+a^2\left(dr^2 +r^2 d\Omega_3{}^2\right).
\label{instanton}
\end{eqnarray}
%
Two 3-branes are located at $y=0$ and $y_0$, as shown in Fig. 2.
Since both $S_E$ and $S_E^{(0)}$ diverge, we introduce
a cutoff at $r=r_c$ such that $S_E-S_E^{(0)}$ is well-defined
in the limit  $r_c \to +\infty$.

The Euclidean action of the system is given by
%
\begin{eqnarray}
S_E=S_{\rm bulk}+S_{\rm brane}+S_{\infty}+S_p,\label{contributions}
\end{eqnarray}
%
where
%
\begin{eqnarray}
S_{\rm bulk}&=&\int_{M}\sqrt{g}\left(\Lambda-\frac{1}{2}R\right)d^5 x,\\
S_{\rm brane}&=&\int_{B}\sqrt{^{(4)}g}\frac{\lambda_+}{2}d^4 x
-\int_{B}\sqrt{^{(4)}g}Kd^4 x \nonumber \\
& & +\int_{\tilde B}\sqrt{^{(4)}g}\frac{\lambda_-}{2} d^4x
-\int_{\tilde B}\sqrt{^{(4)}g}Kd^4 x, \\
S_{\infty}&=&-\int_{r=r_c}\sqrt{^{(4)}g}Kd^4 x,\\
S_p&=&-\int_{p}\sqrt{^{(4)}g}Kd^4 x.
\end{eqnarray}
%
Here, the positive and negative tension branes are
denoted by $B$ and $\tilde B$, respectively,
$\lambda_\pm=\pm 6/\ell$ is the tension of each brane,
$\Lambda=-6/\ell^2$ is the bulk cosmological constant,
and $M$ is the bulk with the cutoff $r=r_c$. 
The quantity $S_p$ is the contribution from the edge $p$,
which is the surface of intersection of $B$ and $\tilde B$.
Since there is a deficit angle, the introduction 
of $S_p$ corresponds to that of a string-like defect 
at the intersection.
\footnote{
There is an ambiguity with regard to how one treats this term.
This will depend on the microprocesses when two branes collide.
We consider a simple case and
leave this problem for future investigations.}

{}From the Einstein equation and the junction condition under $Z_2$ symmetry,
%
\begin{eqnarray}
S_{\rm bulk}=\frac{4}{\ell^2}Vol_5(M)\label{volume}
\end{eqnarray}
%
and
%
\begin{eqnarray}
S_{\rm branes}&=&-\frac{1}{\ell}\left[Vol_4(B)-Vol_4(\tilde{B})\right]
\end{eqnarray}
%
are obtained, 
where $Vol_n$ denotes the $n$-dimensional volume.

\subsection{Euclidean action of the RS vacuum}

For the RS vacuum the volumes of the bulk and the branes become
%
\begin{eqnarray}
Vol_5(M)=
2\pi^2 \int^{y_{0}}_{0}dy \int^{r_c}_{0}dr a^4 r^3=\frac{\pi^2}{8}\left(1-e^{-4y_0/\ell}\right)\ell r_c^4,
\end{eqnarray}
%
%
\begin{eqnarray}
Vol_4(B)=2\pi^2 \int^{r_c}_0 dr r^3 a^4=\frac{\pi^2}{2}r_c^4,
\end{eqnarray}
%
%
\begin{eqnarray}
Vol_4(\tilde{B})=2\pi^2\int^{r_c}_0 dr r^3 a^4
=\frac{\pi^2}{2} r_c^4 e^{-4 y_0/\ell}.
\end{eqnarray}
%
Each contribution to the action therefore becomes
%
\begin{eqnarray}
S_{\rm bulk}^{(0)}&=&\frac{1}{2}\pi^2\left(1-e^{-4y_0/\ell}\right)\frac{r_c^4}{\ell},\\
S_{\rm brane}^{(0)}&=&-\frac{1}{2}\pi^2 \left(1-e^{-4y_0/\ell}\right)\frac{r_c^4}{\ell},\\
S_{\infty}^{(0)}&=&-3\pi^2 \left(1-e^{-2y_0/\ell}\right)\ell r_c^2.
\end{eqnarray}
%
Since there are no edges in the RS vacuum, $S_p^{(0)}=0$. Thus 
the total Euclidean action of the RS vacuum is given by 
%
\begin{eqnarray}
S_E^{(0)}=S^{(0)}_\infty=-3\pi^2 \ell r_c^2\left(1-e^{-2y_0/\ell}\right).
\end{eqnarray}
%
Note that the contributions from the bulk and the brane exactly cancel,
due to the exact balance of the bulk cosmological constant and the tension of the flat branes:
%
\begin{eqnarray}
S_{\rm bulk}^{(0)}+S_{\rm brane}^{(0)}=0.
\end{eqnarray}
%

\subsection{The Euclidean action of an RS bubble}

For the RS bubble space-time, the volumes of the bulk and branes are 
%
\begin{eqnarray}
Vol_5(M) &=&2\pi^2\int^{y_{0}}_{0}dy \int^{r_c}_{r_0(y)}dr a^4 r^3
\Bigl[1-\Bigl(\frac{\rho_*}{ar}\Bigr)^2\Bigr] 
\biggl[1+\Bigl(\frac{\rho_*}{ar}\Bigr)^2\biggr]^3 \nonumber \\ 
&=& \pi^2 \rho_*^4 \ell \biggl[\frac{1}{8}\Bigl( \frac{\rho_*}{r_c}\Bigr)^4
(a_0^{-4}-1)+\Bigl( \frac{\rho_*}{r_c}\Bigr)^2(a_0^{-2}-1)\nonumber \\
& & -\Bigl(\frac{r_c}{\rho_*} \Bigr)^2(a_0^2-1)
-\frac{1}{8}\Bigl( \frac{r_c}{\rho_*}\Bigr)^4(a_0^4-1) \biggr] \nonumber \\
& &-\pi^2\rho_*^4 \int^{y_0}_0 dy \biggl[\frac{1}{2}
\left(\frac{a r_0(y)}{\rho_*}\right)^{-4} 
+2\left(\frac{a r_0(y)}{\rho_*}\right)^{-2} \nonumber\\
& &+2 \left(\frac{a r_0(y)}{\rho_*}\right)^{2} 
+\frac{1}{2}\left(\frac{a r_0(y)}{\rho_*}\right)^{4} \biggr],
\end{eqnarray}
%
%
\begin{eqnarray}
Vol_4(B)&=&2\pi^2\int_{r_0(0)}^{r_c}r^3\left[1+\left(\frac{\rho_*}{r}\right)^2\right]^4dr
\nonumber\\
&=&\pi^2 \rho_*^4\left[ 
\frac{1}{2} \left(\frac{r_c}{\rho_*}\right)^4\right.
+4\left( \frac{r_c}{\rho_*}\right)^2
+12 \ln \left(\frac{r_c}{\rho_*}\right)
-4\left(\frac{r_c}{\rho_*}\right)^{-2}
\nonumber \\
&&-\frac{1}{2} \left(\frac{r_c}{\rho_*}\right)^{-4}
-\frac{1}{2}\left(\frac{r_0(0)}{\rho_*} \right)^4
{}-4\left(\frac{r_0(0)}{\rho_*} \right)^2
-12\ln\left(\frac{r_0(0)}{\rho_*}\right)
\nonumber \\
&&\left.+4\left(\frac{r_0(0)}{\rho_*}\right)^{-2}
+\frac{1}{2}\left(\frac{r_0(0)}{\rho_*}\right)^{-4}
\right],
\end{eqnarray}
%
%
\begin{eqnarray}
Vol_4(\tilde{B}) &=&
2\pi^2\int_{\rho_*^2/a_0 r_0(0)}^{r_c}a_0{}^4r^3
\left[1+\left(\frac{\rho_*}{a_0 r}\right)^2\right]^4dr
\nonumber\\
&=&\pi^2 \rho_*^4\left[ 
\frac{1}{2} \left(\frac{a_0 r_c}{\rho_*}\right)^4\right.
+4\left( \frac{a_0 r_c}{\rho_*}\right)^2
+12 \ln \left(\frac{a_0 r_c}{\rho_*}\right)
-4\left(\frac{a_0r_c}{\rho_*}\right)^{-2}
\nonumber \\
&&-\frac{1}{2} \left(\frac{a_0r_c}{\rho_*}\right)^{-4}
-\frac{1}{2}\left(\frac{r_0(0)}{\rho_*} \right)^4
{}-4\left(\frac{r_0(0)}{\rho_*} \right)^2
-12\ln\left(\frac{r_0(0)}{\rho_*}\right)
\nonumber \\
&&\left.+4\left(\frac{r_0(0)}{\rho_*}\right)^{-2}
+\frac{1}{2}\left(\frac{r_0(0)}{\rho_*}\right)^{-4}
\right],
\end{eqnarray}
%
where $a_0=e^{-y_0/\ell}$, and $r_0 (y)$ is 
implicitly defined as the larger solution of the equations
%
\begin{eqnarray}
\frac{y_0-y}{2}&=&\frac{1}{\ell}\int^{R}_{R_*}dR R
\left[1-\left(\frac{2\rho_*}{R}\right)^2
+\left(\frac{R}{\ell}\right)^2\right]^{-1}
\left[1-\left(\frac{2\rho_*}{R}\right)^2\right]^{-1/2},\label{ct1}\\
R&=&a(y) r_0(y)\left[1+\left(\frac{\rho_*}{a(y) r_0(y)}\right)^2\right].
\end{eqnarray}
%
As seen below, we need a numerical computation to determine $r_0(y)$ in general. 

Though both $S_{\rm bulk}$ and $S_{\rm brane}$ are unbounded, the 
divergent terms again cancel, and we obtain 
%
\begin{eqnarray}
S_{\rm bulk}+S_{\rm brane}&=&
\frac{\pi^2\rho_*^4}{\ell}\left\{
\left(\frac{r_0(0)}{\rho_*}\right)^4
+8\left(\frac{r_0(0)}{\rho_*}\right)^2
+24\ln\frac{r_0(0)}{\rho_*}
-8\left(\frac{r_0(0)}{\rho_*}\right)^{-2}
\right.\nonumber\\
&&
{}-\left(\frac{r_0(0)}{\rho_*}\right)^{-4}
-12\frac{y_0}{\ell}
-\frac{1}{\ell}\int_0^{y_0}\left[
2\left(\frac{ar_0}{\rho_*}\right)^4
+8\left(\frac{ar_0}{\rho_*}\right)^2
\right.\nonumber\\
&&
\left.\left.
+8\left(\frac{ar_0}{\rho_*}\right)^{-2}
+2\left(\frac{ar_0}{\rho_*}\right)^{-4}
\right]dy\right\}\nonumber\\
&=&
\frac{\pi^2\rho_*^4}{\ell}\left\{
16\frac{\rho_p^4}{\rho_*^4}\left(1+\frac{3\rho_*^2}{2\rho_p^2}\right)
\sqrt{1-\frac{\rho_*^2}{\rho_p^2}}
+24\ln\frac{\rho_p}{\rho_*}\left(1+\sqrt{1-\frac{\rho_*^2}{\rho_p^2}}\right)
\right. \nonumber\\
&&-12\frac{y_0}{\ell}
-\frac{1}{\ell}\int_0^{y_0}\left[
2\left(\frac{ar_0}{\rho_*}\right)^4
+8\left(\frac{ar_0}{\rho_*}\right)^2
\right.\nonumber\\
&&
\left.\left.
+8\left(\frac{ar_0}{\rho_*}\right)^{-2}
+2\left(\frac{ar_0}{\rho_*}\right)^{-4}
\right]dy\right\}, \label{39}
\end{eqnarray}
%
where
\begin{equation}
\rho_p:=\frac{r_0(0)}{2}\left[1+\left(\frac{\rho_*}{r_0(0)}\right)^2\right]
\end{equation}
denotes half the volume radius at $p$. As seen below 
$S_{\rm bulk}+S_{\rm brane}$ is much smaller than $S_p$. 

The $r=r_c$ term $S_{\infty}$ of the Euclidean action
is estimated as
%
\begin{equation}
S_{\infty}=-3\pi^2r_c^2 \ell(1-e^{-2y_0/\ell})+4\pi^2\rho_*^2y_0. \label{inftyaction}
\end{equation}
%
This shows that the contribution from the boundary is also
finite: $[S_{\infty}]=S_{\infty}-S_{\infty}^{(0)}=4\pi^2\rho_*^2y_0 < +\infty$.

Following Ref. ~\cite{Hay}, $S_p$ is evaluated as  
%
\begin{equation}
S_p=16\pi^2\rho_p^3 \Delta\phi_p,
\end{equation}
%
where $\Delta\phi_p$ is the exterior angle in the ($R$,$\Theta$)-plane at $p$.
The exterior angle $\Delta\phi_p$ is evaluated as 
%
\begin{eqnarray}
\Delta\phi_p & = & {\rm cos}^{-1}(-g_{\mu\nu}n^\mu_+ n^\nu_- )\nonumber \\
& = & \cos^{-1}\left[\frac
{4\left(\rho_p/\ell\right)^2
+\left(\rho_*/\rho_p\right)^2-1}
{4\left(\rho_p/\ell\right)^2
-\left(\rho_*/\rho_p\right)^2+1}\right],
\end{eqnarray}
%
where $n_{\pm}$ are the unit normal vectors of the $(\pm)$-branes at $p$
given by
%
\begin{equation}
n_{\pm}=\sqrt{
1-\frac{\rho_*^2}{\rho_p^2}
}
\left[
-\frac{1}{F(2\rho_p)}\partial_{\Theta}
\pm
\frac{2\rho_p}{\ell\sqrt{
1-\rho_*^2/\rho_p^2}
}
\partial_R
\right].
\end{equation}
%
Thus we obtain
%
\begin{equation}
S_p=16\pi^2 \rho_p^3
\cos^{-1}\left[\frac
{4\left(\rho_p/\ell\right)^2
+\left(\rho_*/\rho_p\right)^2-1}
{4\left(\rho_p/\ell\right)^2
-\left(\rho_*/\rho_p\right)^2+1}\right].\label{paction}
\end{equation}
%
Gathering Eqs. (\ref{39}) (\ref{inftyaction}) and (\ref{paction}), we 
obtain the total action, Eq. (\ref{contributions}). 

\subsection{Estimation of the decay rate}

\begin{figure}[t]
\vspace*{-5mm}
\begin{center}
\epsfxsize=3.0in
\epsffile{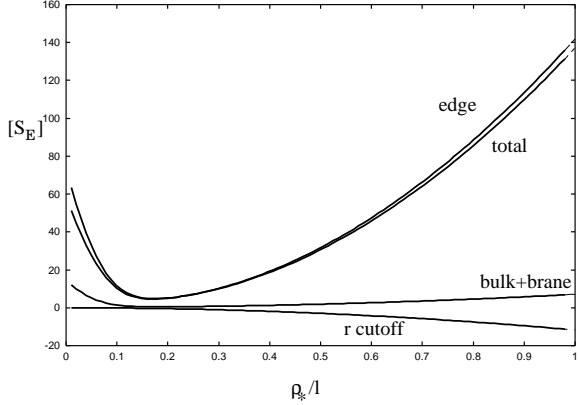}
\end{center}
\caption{The Euclidean action between the
RS vacuum and the bubble for $y_0=\ell=1$.
Here ``bulk+brane'', ``$r$ cutoff'', ``edge'' and ``total'' correspond to
$S_{\rm bulk}+S_{\rm brane}$, $S_{\infty}$, $S_{p}$ and $[S_{E}]$,
respectively.}
\label{fig4}
\end{figure}

\begin{figure}[t]
\vspace*{-5mm}
\begin{center}
\epsfxsize=3.0in
\epsffile{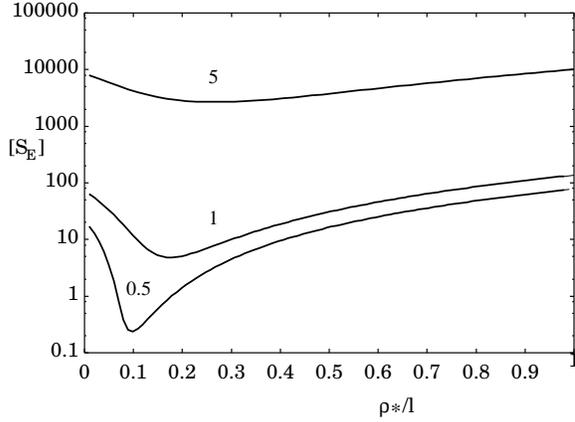}
\end{center}
\caption{$[S_E]$ for $\ell=1$ and $y_0=0.5,1,5$. In all cases $S_p$ dominates.}
\label{fig5}
\end{figure}

\begin{figure}[t]
\vspace*{-5mm}
\begin{center}
\epsfxsize=3.0in
\epsffile{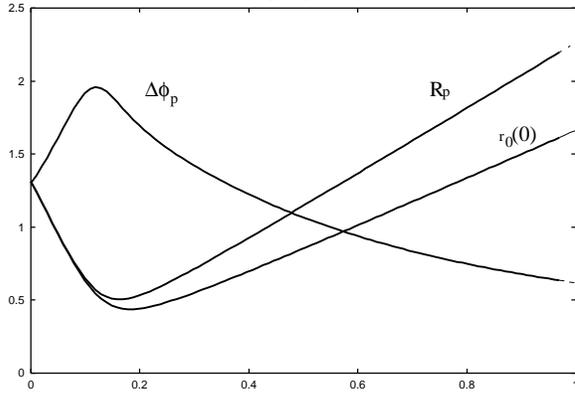}
\end{center}
\caption{$R_p$, $r_0(0)$ and $\Delta \phi_p$ for $\ell=1$. 
$R_p$ determines the behavior of $S_p$.}
\label{fig6}
\end{figure}

We are now ready to evaluate the transition probability from the RS 
vacuum to the RS bubble,
%
\begin{equation}
P\sim \exp\left(-[S_E]\right).
\end{equation}
%
In general, we need a numerical computation to evaluate $[S_E]$.
The numerical results are 
given in Figs.~\ref{fig4} and \ref{fig5}.
It is seen that the dominant contribution to $[S_E]$ 
comes from $S_p$,
while the contribution from $S_{\rm brane}+S_{\rm bulk}$ is relatively small. 
In addition we can see from Fig. \ref{fig6} that the $\rho_*$ dependence 
of  $S_p$ is mainly governed by $R_p$. 
The contribution $S_p$ is a non-monotonic function at small $R_*$. 
The volume effect of RS bubble
reduces the decay rate for relatively large $\rho_*$. 
In any case, an RS bubble with small $\rho_*$ and small $y_0$ 
will be nucleated easily. 
Once created, a bubble quickly expands at nearly the speed of light and 
eventually occupies the entire universe. For $y_0 \sim 40 \ell$, however, 
we can see from Fig. 5 that the transition amplitude is significantly 
suppressed. 

Finally, we give some useful analytic expressions using some limiting conditions. 
When $R_*=0$, 
the function $r_0(y)$ is given by
\begin{equation}
a(y)r_0(y)=\ell \sqrt{ e^{(y_0-y)/\ell}-1},
\end{equation}
and $\rho_p$ becomes
\begin{equation}
\rho_p=\frac{\ell}{2} \sqrt{e^{y_0/\ell}-1}.
\end{equation}
Then, we obtain
\begin{eqnarray}
&&S_{\rm bulk}+S_{\rm brane} = 2\pi^2 \ell^3
\left[e^{y_0/\ell}-1-\frac{y_0}{\ell} \right],\\
& & S_{\infty}-S^{(0)}_{\infty}  =0,\\
& & S_p=16\pi^2\rho_p^3\cos^{-1}\left[\frac
{4\left(\rho_p/\ell\right)^2-1}
{4\left(\rho_p/\ell\right)^2+1}
\right].
\end{eqnarray}
With the further limiting condition $y_0/\ell \ll 1$,
the difference from the Euclidean action becomes
\begin{equation}
[S_E]\simeq S_p \simeq \frac{\pi^2 \ell^3}{G_5} \Bigl(\frac{y_0}{\ell} 
\Bigr)^{3/2},
\end{equation}
where  $G_5$ has been recovered. 

When $R_*/\ell \gg 1$ on the other hand, we have
$1-(R_*/R)^2\ll (R/\ell)^2$, so that  
the coordinate transformation can be explicitly
carried out in this case.  Then, $r_0(y)$ is simply given by 
%
\begin{equation}
a(y)r_0(y) \simeq \rho_*\exp\left(\pm\frac{y_1-y}{2\ell}\right),
\end{equation}
%
where the upper sign is for the bulk and the positive tension brane, 
and the lower sign is for the negative tension brane.
In this limit, we can see that the term $S_{\rm bulk}+S_{\rm brane}$
has only a small contribution. In addition, we have
$[S_\infty]\simeq -6\pi^2\rho_*^2 \ell 
(1-e^{-2y_0/\ell})+4\pi^2\rho_*^2 y_0$ and $S_p \simeq 8\pi^2\rho_*^2 \ell 
\sinh( y_0 / \ell )$. Then $[S_E]$ becomes 
%
\begin{eqnarray}
[S_E]& \simeq & [S_{\infty}]+S_p \nonumber \\
& \simeq &  -6\pi^2\rho_*^2 \ell 
(1-e^{-2y_0/\ell})+4\pi^2\rho_*^2 y_0\nonumber\\
& &+8\pi^2\rho_*^2 \ell \sinh( y_0 / \ell ). \label{appro}
\end{eqnarray}
%
Note that the bulk term and the brane term of the Euclidean action 
cancel out.
With the further condition $y_0/\ell \ll 1$,
the difference of the Euclidean action is
%
\begin{eqnarray}
[S_E]\simeq 48\pi^2\frac{\rho_*^2 y_0^2}{G_5 \ell},
\end{eqnarray}
%
while for $y_0 \sim \ell $, we have
%
\begin{eqnarray}
[S_E]\simeq 8 \pi^2 \frac{\rho_*^2 \ell}{G_5}.
\end{eqnarray}
%
If we are interested in the hierarchy problem, we can
set $y_0 \sim 40 \ell$\cite{RSI} in Eq. (\ref{appro}).
Then we obtain
%
\begin{eqnarray}
[S_E] \simeq 10^{19} \times 
\Bigl( \frac{\rho_*^2 \ell}{G_5} \Bigr). 
\label{S-hier}
\end{eqnarray}
%
Except for the large numerical prefactor, the dependence of 
$[S_E]$ on the physical parameters $\rho_*$ and $\ell$ is 
found in a previous paper.~\cite{rsb}
It should be noted that 
we cannot set $\rho_*=0$ in Eq.~(\ref{S-hier}) because we are assuming
$\rho_* / \ell \gg 1$.

\section{Summary}\label{IV}
Let us summarize our study. We have investigated the instanton
solution that describes the decay of the RS vacuum into an RS bubble. 
The decay probability is numerically estimated.
We also derived an analytic formula for the decay probability into
large RS bubbles.
We found that the RS vacuum (RS1 model) is unstable in general.
We also found that decay into a small RS bubble is 
favored over decay into a large RS bubble, as be expected. 
However, the transition rate is suppressed 
when two branes are sufficiently separated as is assumed in the
context of the hierarchy problem. 

Our result brings to light a problem concerning the stability of the 
brane world. 
The entire space will be quickly occupied by RS bubbles.
This is, however, a topology-changing process,
so that this instability might be forbidden by the introduction of spinor fields. 
A supersymmetry might also serve as a stabilizer.

\section*{Acknowledgements}
HO would like to thank K. Sato for his continuous encouragement. 
TS's work is partially supported by the Yamada Science Foundation.


\begin{thebibliography}{99}
\bibitem{HW}
P. Horava and E. Witten, Nucl. Phys. {\bf B460}(1996),506;
{\bf B475} (1996), 94.
\bibitem{RSI}
L.~Randall and R.~Sundrum, Phys. Rev. Lett. {\bf 83}(1999), 3370. 
\bibitem{RSII}
L.~Randall and R.~Sundrum, Phys. Rev. Lett. {\bf 83}(1999), 4690
\bibitem{Tama}
J.~Garriga and T.~Tanaka, Phys. Rev. Lett. {\bf 84} (2000), 2778;\\
S.~B.~Giddings, E.~Katz, and L.~Randall, JHEP {\bf 0003}(2000), 023.
\bibitem{Tess} 
T.~Shiromizu, K.~Maeda, M.~Sasaki, Phys. Rev. {\bf D62} (2000), 024012;\\
M.~Sasaki, T.~Shiromizu and K.~Maeda, Phys. Rev. {\bf D62}, (2000), 024008.
\bibitem{Cosmos}
P.~Bin\'{e}truy, C.~Deffayet, U.~Ellwanger and D.~Langlois, Phys. Lett.
{\bf B477}(2000), 285.\\
P.~Kraus, JHEP {\bf 9912}(1999), 011.\\
D.~Ida, JHEP {\bf 0009}(2000), 014.\\
S.~Mukohyama, Phys. Lett. {\bf B473} (2000), 241.\\
S.~Mukohyama, T.~Shiromizu and K.~Maeda, Phys. Rev. {\bf D62} (2000), 024028.
\bibitem{adSCFT}
S. S. Gubser, Phys. Rev. {\bf D63} (2001), 084017.\\ 
M.J. Duff and J. T. Liu, Phys. Rev. Lett. {\bf 85} (2000), 2052.\\
M. Perez-Victoria, hep-th/0105048.\\
T. Shiromizu and D. Ida, Phy. Rev. {\bf D64} (2001), 044015.\\
A. Hebecker and J. March-Russell, hep-ph/0103214.\\
S. W. Hawking, T. Hertog and H. S. Reall, Phys. Rev. {\bf D62}(2000), 
043501.\\ 
L. Anchordoqui, C. Nunez and K. Olsen, JHEP {\bf 0010}(2000), 050.\\
S. Nojiri, S.D. Odintsov and S. Zerbini, 
Phys. Rev. {\bf D62} (2000), 064006.\\
S. Nojiri and S.D. Odintsov, Phys. Lett. {\bf B484} (2000), 119.
\bibitem{Tama2}
T.~Tanaka and X. Montes, Nucl. Phys. {\bf B582}(2000), 259.
\bibitem{gw}
W. D. Goldberger and M. B. Wise,
Phys. Rev. Lett. {\bf 83} (1999), 4922.
\bibitem{KKB}
E.~Witten, Nucl. Phys. {\bf B195}(1982), 481.
\bibitem{Gary}
D.~Brill and H.~Pfisher, Phys. Lett. {\bf B228}(1989), 359.\\
D.~Brill and G.~T.~Horowitz, Phys. Lett. {\bf B262}(1991), 437.
\bibitem{Shinkai}
H.~Shinkai and T.~Shiromizu, Phys. Rev. {\bf D62}(2000), 024010.
\bibitem{Horava}
M.~Fabinger and P.~Horava, Nucl. Phys. {\bf B580}(2000), 243.
\bibitem{rsb}
D. Ida, T. Shiromizu and H. Ochiai, Phys. Rev. {\bf D65} (2002) 023504.
\bibitem{Ruth}
R. Gregory and A. Padilla, hep-th/0104262; hep-th/0107108.
\bibitem{dSCFT}
T.~Shiromizu, D.~Ida and T.~Torii, JHEP {\bf 0111} (2001) 010..
\bibitem{Iso}
G. C. MacVittie, Month. Not. Roy. Astron. Soc. {\bf 93}(1933), 325.\\
M. Kihara and H. Nariai, Prog. Theor. Phys. {\bf 65}(1981), 1613.
\bibitem{ER}
A.~Einstein and N.~Rosen, Phys. Rev. {\bf 48}(1935), 73.
\bibitem{Gen}
R.~Bousso and A.~Chamblin, Phys. Rev. {\bf D59}(1999), 084004.\\
U.~Gen and M.~Sasaki, Phys. Rev. {\bf D61}(2000), 103508.
\bibitem{Hay}
G.~Hayward, Phys. Rev. {\bf D47}(1993), 3275.

\end{thebibliography}
\end{document}